# Microwave plasma CVD growth and characterization of polycrystalline diamond films on β-$Ga_2O_3$

Saleh Ahmed Khan[1], Stephen Margiotta[1], Ahmed Ibreljic[1], Anhar Bhuiyan[1, a)]

[1]*Department of Electrical and Computer Engineering, University of Massachusetts Lowell, MA 01854, USA*

[a)] Corresponding author Email: anhar_bhuiyan@uml.edu

**Abstract**

The integration of diamond with β-$Ga_2O_3$ presents a promising pathway to enhance thermal management in high-power electronic devices, where the inherently low thermal conductivity of β-$Ga_2O_3$ can lead to localized self-heating and elevated junction temperatures. In this work, we demonstrate a scalable, low-damage approach to integrate polycrystalline diamond films on (010) β-$Ga_2O_3$ substrates via microwave plasma chemical vapor deposition (MPCVD), employing dielectric interlayers and polymer-assisted electrostatic nanodiamond seeding to systematically evaluate the impact of growth conditions on film morphology, grain evolution, phase purity, and optical characteristics. At the growth temperature of 800 °C, progressive grain coarsening is observed with extended deposition, with lateral grain size increasing from 37.6 nm (53 nm thickness) to 192.5 nm for an 886 nm film. This microstructural evolution is accompanied by narrowing of the diamond Raman peak and a monotonic increase in $sp^3$-phase fraction from 95.9% to as high as 98.9%, indicating continued suppression of non-diamond carbon with prolonged growth. Comparison of $SiO_2$ and $SiN_x$ interlayers under identical growth conditions shows only marginal differences in grain size and phase purity, indicating limited interlayer influence once high nucleation density is established. Importantly, diamond films exhibiting >96% $sp^3$-phase content were achieved at substrate temperatures as low as 480 °C, highlighting the viability of diamond-on-$Ga_2O_3$ integration under reduced thermal budgets. These findings establish a robust, scalable platform for integrating diamond on β-$Ga_2O_3$, supporting the development of next-generation power and RF devices with improved thermal management.

The ultrawide bandgap β-$Ga_2O_3$ has emerged as a promising semiconductor for next-generation power electronics, owing to its large bandgap (~4.8 eV), high critical electric field (~8 MV/cm), and the availability of low-cost, melt-grown large area native substrates [1-4]. However, a critical limitation hindering the widespread adoption of $Ga_2O_3$ in high-power applications is its intrinsically low thermal conductivity (10-27 W/m·K) [5]. This poor heat dissipation capability leads to severe thermal management challenges under high power densities, resulting in localized hot spots, device degradation, and long-term reliability concerns [6,7].

An effective approach to overcoming thermal limitations in β-$Ga_2O_3$ devices involves integrating materials with superior thermal conductivity. Diamond, in particular, offers a compelling solution due to its exceptional thermal conductivity, surpassing 2000 W/m·K in single-crystal form [8] and reaching 1000-1800 W/m·K in optimized polycrystalline films [9], along with its wide bandgap, outstanding mechanical strength, and chemical stability [10-13]. Prior efforts involving GaN-on-diamond integration have demonstrated substantial reductions in thermal resistance and junction temperature through the use of microwave plasma-enhanced chemical vapor deposition (MPCVD)-grown polycrystalline diamond layers. These advances were enabled by careful interface engineering to promote nucleation and protect the underlying device structure during growth [14-23]. Various seeding strategies have also been developed to address the challenge of initiating diamond nucleation on non-diamond surfaces, including mechanical abrasion [24-27], bias-enhanced nucleation [28, 29], chemical nucleation using adamantane derivatives [30-33], and polymer-assisted electrostatic methods [34-36]. These approaches have achieved nucleation densities up to ~$10^{11}$ $cm^{-2}$ [37-38] and enabled coalesced films on foreign substrates.

However, translating this success to β-$Ga_2O_3$ has proven non-trivial. The surface energy mismatch between diamond (~6 J/m²) and β-$Ga_2O_3$ (~1 J/m²) can suppress nucleation and hinder lateral growth, often leading to sparse, discontinuous films or isolated diamond clusters unless appropriate interface treatments are applied [39-41]. Although a few studies have demonstrated the feasibility of diamond growth on β-$Ga_2O_3$ using interlayers such as $SiO_2$ [42] or $Al_2O_3/SiO_2$ [39] in combination with nanodiamond seeding [39,42,43], these efforts have primarily emphasized early-stage nucleation behavior. Systematic studies that explore how downstream MPCVD growth conditions affect film morphology, structural ordering, phase composition and optical characteristics remain largely unexplored. A deeper understanding of these interdependencies is essential to guide process optimization, improve material quality, and enable reproducible integration of diamond layers on β-$Ga_2O_3$. In this work, we conduct a systematic study of polycrystalline diamond growth on β-$Ga_2O_3$ substrates using MPCVD, targeting the development of conformal diamond films suitable for monolithic integration. By employing dielectric interlayers, followed by polymer-assisted electrostatic seeding with nanodiamond particles, we explore how growth temperature, interfacial chemistry, and deposition time influence film morphology, grain size, surface roughness, phase composition, and optical response. This comprehensive approach offers new insight into how MPCVD process conditions influence the growth kinetics, microstructure, and quality of diamond films on $Ga_2O_3$, providing a foundation for future development of thermal management solutions in high-power electronics.

The diamond films were deposited on (010) β-$Ga_2O_3$ substrates using MPCVD method. To facilitate heterogeneous nucleation of diamond and simultaneously mitigate plasma-induced damage at the $Ga_2O_3$-diamond interface, a dielectric buffer layer with a thickness of ~50 nm was deposited via plasma-enhanced chemical vapor deposition at 300 °C. Two types of dielectric materials, amorphous $SiO_2$ and $SiN_x$, were explored to evaluate their influence on seeding efficiency and film morphology. To ensure uniform and high-density nucleation of diamond, we employed a three-step polymer-assisted nanodiamond (ND) seeding process, leveraging electrostatic self-assembly principles [44]. Each dielectric-coated $Ga_2O_3$ sample was immersed for 10 minutes in an aqueous solution of poly-diallyldimethylammonium chloride (PDDAC), a high molecular weight cationic polymer that renders the surface positively charged through quaternary ammonium functional groups [21,45]. Following this, the sample was immersed in a suspension of ND particles with diameters ranging from 5-10 nm and zeta potentials exceeding ±50 mV. The PDDAC and ND dipping steps were repeated three times to form a trilayer structure, ensuring a dense and uniform coverage of nanodiamond particles across the substrate. Subsequent rinsing with DI water and nitrogen blow drying removed unbound particles. The positive surface potential imparted by PDDAC enhances electrostatic attraction with negatively charged nanodiamonds [44], facilitating high nucleation density and uniform diamond film coalescence during growth. Diamond film growth was performed in a Seki Technotron AX5010-INT MPCVD tool. A hydrogen-rich $CH_4/H_2$ (3/300) mixture containing 1% methane was used as the plasma chemistry, with a total chamber pressure of 50 Torr and a constant microwave power of 1000 W. The substrate temperature ranged from 480 to 800°C. Growth durations were tuned from 10-150 minutes. In addition to β-$Ga_2O_3$, control growth experiments were carried out on sapphire and Si substrates using the same polymer-assisted nanodiamond seeding protocol and comparable MPCVD conditions. On sapphire, films up to ~824 nm thickness exhibit $sp^3$-phase purity of ~98.5%, while on Si substrates, films up to ~814 nm thickness reach ~98.9% $sp^3$-phase purity. The corresponding data are provided in Figures S1 and S2 of the Supplementary Material. Post-growth characterization was conducted to evaluate structural, morphological, and optical properties. Surface morphology, grain size, and film continuity were examined using JEOL JSM 7401F field-emission scanning electron microscope (FESEM) while Raman spectra were recorded with a Horiba LabRam Evolution Multiline Raman Spectrometer employing a 532 nm diode-pumped solid-state laser. Surface roughness and topography were analyzed by atomic force microscopy (AFM) using a Park XE-10, and optical absorption characteristics were obtained from room-temperature UV-Vis-NIR spectrophotometry by utilizing an Agilent Cary 60 UV/Vis spectrophotometer.

Figure 1 presents the evolution of surface morphology and roughness of diamond films deposited on $SiO_2$-coated β-$Ga_2O_3$ at a fixed substrate temperature of 800 °C. Growth durations were varied from 10 to 150 minutes to investigate grain evolution and faceting behavior under low-methane and hydrogen-rich plasma conditions. FESEM images [Figures 1(a-d)] show a progressive transition from densely packed nanocrystals to increasingly faceted and coarsened grain structures with increasing deposition time and thickness. After 10 minutes [Figure 1(a)], the film is composed of densely packed, rounded grains with an average lateral grain size of 37.6 nm and film thickness of ~53 nm. Grain boundaries are not well defined, indicating early-stage vertical growth with limited lateral expansion. This morphology is consistent with high nucleation density achieved through multilayer polymer-assisted seeding using PDDAC and nanodiamond particles, which promotes rapid surface coverage and suppresses secondary nucleation [42]. At 30 minutes [Figure 1(b)], the grain size increases significantly to 91.0 nm, and faceting becomes more pronounced, suggesting the onset of competitive grain growth. After 60 minutes of deposition [Figure 1(c)], the grain size further increases to 126.6 nm with well-defined grain boundaries and angular crystallites, indicating that lateral growth and facet sharpening dominate in the steady-state growth regime. The corresponding film thickness increases to 315 nm, confirming a vertical deposition rate of 5.25 nm/min. Upon extending the growth duration to 150 minutes [Figure 1(d)], the film thickness increases to 886 nm (growth rate of 5.91 nm/min), and the average lateral grain size further increases to 192.5 nm. The microstructure exhibits well-developed faceted grains with reduced grain boundary density, indicative of continued grain competition and coalescence. This progressive grain coarsening with thickness is consistent with thickness-dependent microstructural refinement during prolonged MPCVD growth and aligns with grain selection and competitive faceting behavior reported for low-$CH_4$ diamond growth on GaN and oxide-buffered substrates [18,21,42]. AFM topography scans, shown in Figures 1(e-h), quantitatively confirm the systematic increase in surface roughness associated with grain coarsening. The root mean square (RMS) roughness increases from 9.4 nm at 10 minutes to 13.2 nm at 30 minutes, reaching 16.3 nm at 60 minutes, and further increasing to 25.7 nm at 150 minutes. This monotonic increase reflects enhanced facet development and increased vertical height contrast between coalesced grains as the film thickens. Importantly, despite the substantial increase in thickness to nearly 900 nm, the film remains continuous, conformal, and free of delamination or pinholes, demonstrating robust adhesion to the underlying $SiO_2$ interlayer. The $SiO_2$ serves as a thermally stable and chemically inert diffusion barrier, while also providing a negatively charged surface conducive to electrostatic nanodiamond seed attachment [39,42].

Figure 2 shows the room-temperature Raman spectra of the diamond films grown on $SiO_2$/β-$Ga_2O_3$ substrates at 800 °C for deposition times ranging from 10 to 150 minutes. All spectra display the first-order Raman peak associated with $sp^3$-bonded diamond near 1332 $cm^{-1}$, confirming successful phase formation under hydrogen-rich MPCVD conditions. A systematic evolution of the diamond peak is observed with increasing growth duration and film thickness. The full width at half maximum (FWHM) decreases from 74.6 $cm^{-1}$ for the 53 nm film to 65.5 $cm^{-1}$ for the 157 nm film and 52.98 $cm^{-1}$ for the 315 nm film, reaching 51.34 $cm^{-1}$ for the 886 nm thick film. This progressive narrowing is consistent with the grain coarsening observed in FESEM and indicates improved structural order and reduced phonon confinement as the microstructure evolves [21,46]. In addition to peak sharpening, the relative intensity of the non-diamond carbon features decreases with continued growth. Secondary spectral features in the 1100-1600 $cm^{-1}$ region include a shoulder near ~1150 $cm^{-1}$ attributed to trans-polyacetylene and a G-band centered around ~1550 $cm^{-1}$ associated with disordered $sp^2$-bonded carbon [21,42,43]. While these features are more pronounced in thinner films, their relative contribution diminishes in thicker films, indicating progressive suppression of $sp^2$-bonded carbon with extended deposition.

Figure 3 presents the evolution of key morphological and spectroscopic properties as a function of film thickness. Data points correspond to the four films characterized in Figures 1 and 2, with measured thicknesses of 53, 157, 315, and 886 nm for growth durations of 10, 30, 60, and 150 minutes, respectively. In Figure 3(a), both average grain size and surface roughness show a clear and systematic increase with increasing film thickness. These trends confirm that grain coarsening and surface topography evolve concurrently as the film develops. The continuous increase in grain size across the entire thickness range indicates sustained competitive grain growth and lateral expansion without significant secondary nucleation. Similar microstructural evolution has been reported in MPCVD-grown diamond, where continued deposition promotes coalescence of favorably oriented crystallites and reduction of grain boundary density [18]. Figure 3(b) shows the corresponding evolution of Raman FWHM and $sp^3$-phase purity. The FWHM of the diamond peak decreases from 74.6 $cm^{-1}$ for the 53 nm film to 51.34 $cm^{-1}$ for the 886 nm thick film. The phase purity of the diamond layer, expressed as the $sp^3$-phase fraction, was estimated from integrated Raman peak area analysis [18,47-49]. In this method, the integrated area of the diamond peak near 1332 $cm^{-1}$ is compared to the combined area of non-diamond carbon contributions, and a correction factor of 75 is applied to account for the higher Raman scattering efficiency of $sp^2$-bonded carbon relative to $sp^3$-bonded carbon [18,48,49]. Using this approach, the estimated $sp^3$-phase purity increases monotonically from 95.9% (53 nm) to 96.7% (157 nm), then to 97.1% (315 nm), and ultimately reaches as high as 98.9% for the 886 nm film. This monotonic enhancement suggests effective suppression of $sp^2$-bonded carbon during extended growth [18,50], as well as better-defined grain boundaries with reduced disorder.

Figure 4 compares the surface morphology and roughness of diamond films grown at three different substrate temperatures: 480 °C, 600 °C, and 800 °C. The growth durations were 120, 60, and 150 minutes, respectively, yielding thicknesses of 58.1 nm, 110.4 nm, and 886 nm. These correspond to growth rates of 29.1, 110.4, and 354.4 nm/hr, respectively. At 480 °C [Figures 4(a-b)], the film exhibits a compact but fine-grained morphology, with an average grain size of 46.8 nm and an RMS roughness of 10.4 nm. The relatively low growth rate and limited adatom mobility at this temperature restrict lateral grain coalescence and facet development, resulting in rounded grains with less distinct boundary definition. As the substrate temperature increases to 600 °C [Figures 4(c–d)], the grain size increases, indicating enhanced surface diffusion. Grain boundaries become sharper, and shallow faceting begins to emerge, consistent with a transition from nucleation-dominated growth to a competitive grain selection regime [18]. The film grown at 800 °C [Figures 4(e-f)] exhibits pronounced grain coarsening and faceted surface morphology, with an average grain size of 192.5 nm and an RMS roughness of 25.7 nm. The increase in growth rate and surface mobility promotes the expansion of energetically favorable crystallites, resulting in polygonal faceting and elevated topographical features. This trend mirrors the morphology and coarsening behavior observed in Figure 1. The higher roughness at this temperature reflects greater vertical height contrast between facets, a signature of later-stage competitive grain growth in hydrogen-rich diamond CVD systems [18, 51]. Across all temperatures, the films remain continuous and conformal without delamination.

Figure 5 presents the room-temperature Raman spectra and optical absorption data for diamond films grown at 480 °C, 600 °C, and 800 °C. In the Raman spectra [Figure 5(a)], each diamond-coated sample exhibits a distinct $sp^3$-bonded diamond peak near 1332 $cm^{-1}$, confirming successful diamond phase formation at all temperatures. The sharpness of this peak increases with growth temperature. This progressive narrowing indicates enhanced crystalline quality, aligning with the grain growth and surface evolution trends observed in SEM and AFM analyses in Figure 4. Because the deposition temperature directly influences the growth rate, the films grown at 480 °C, 600 °C, and 800 °C reached different thicknesses (58 nm, 110 nm, and 886 nm, respectively). Therefore, the observed Raman evolution reflects changes in growth kinetics with temperature and the corresponding differences in film thickness. The appearance of $\beta$-$Ga_2O_3$ substrate phonon modes in all spectra confirms the transparency of the diamond layers and preservation of substrate crystallinity throughout the CVD process. Figure 5(b) shows the optical absorption spectra of the same films, with the inset presenting Tauc plots used to estimate the effective optical absorption edge [52]. A systematic shift of the absorption edge toward higher energies is observed, with extracted values of 3.89 eV (480 °C), 4.22 eV (600 °C), and 5.13 eV (800 °C). This shift correlates with the increase in grain size and $sp^3$-phase purity, and is influenced by both temperature and film thickness, consistent with reduced $sp^2$-related states and progressive microstructural refinement [53].

Figure 6 shows the surface morphological and spectroscopic properties of the films as a function of growth rate, which increases with substrate temperature from 480 °C to 800 °C. It should be noted that the temperature-dependent trends discussed in Figures 4-6 are inherently coupled with variations in growth rate and resulting film thickness; therefore, the observed evolution reflects the combined influence of temperature, growth kinetics, and thickness, rather than an isolated temperature effect. In Figure 6(a), both average grain size and RMS surface roughness increase with increasing growth rate and thickness (from 58 nm to 886 nm). These trends are consistent with a thermally activated transition from compact nanocrystallites at lower temperatures to larger, faceted grains at higher temperatures and greater deposition thickness. The observed morphological evolution is attributed to enhanced surface diffusion and increased adatom mobility, which promote lateral grain competition and selective expansion of energetically favorable crystallites [18,51]. Figure 6(b) shows the corresponding evolution of $sp^3$-phase purity and Raman FWHM. As the growth rate increases from 29.1 nm/hr at 480 °C (58 nm thickness) to 354.4 nm/hr at 800 °C (886 nm thickness), the Raman FWHM narrows from 92.2 $cm^{-1}$ to 51.34 $cm^{-1}$, consistent with improved structural coherence in larger grains [46]. In parallel, the $sp^3$-phase purity increases from 96.1% at 58 nm to 98.9% at 886 nm, reflecting progressive suppression of $sp^2$-bonded carbon during continued growth [18,50]. To evaluate the influence of dielectric interlayers under identical growth conditions, additional comparative Raman, SEM, and optical absorption analyses were performed for diamond films grown on 50 nm $SiO_2$- and $SiN_x$-coated β-$Ga_2O_3$ substrates. The $SiO_2$ sample exhibited an average grain size of 126.6 nm and an $sp^3$-phase purity of 97.1%, whereas the $SiN_x$ sample showed a grain size of 106.5 nm and a phase purity of 96.9%. The extracted effective optical absorption edges were 5.13 eV and 5.11 eV, respectively. These small differences indicate only marginal interlayer influence under identical growth conditions. The complete datasets and detailed discussion are provided in the Supplementary Material (Fig. S3).

Figure 7 provides a comparative benchmarking analysis of $sp^3$-phase purity as a function of diamond film thickness, incorporating representative literature reports of polycrystalline diamond films grown on foreign substrates [17-19, 38, 41, 54-61]. A general trend is observed across prior studies in which thinner films exhibit lower apparent $sp^3$ fractions, while continued growth and increased thickness correspond to progressive suppression of $sp^2$-bonded carbon. The present work follows this established behavior, with $sp^3$-phase purity increasing systematically from ~96% in ultrathin films to values approaching 99% for the 886 nm thick film. Notably, the high-purity data point achieved in this study lies within the upper range of reported values for hetero-integrated diamond films under comparable growth conditions.

In summary, this work presents a comprehensive investigation into the MPCVD growth of diamond films on $\beta$-$Ga_2O_3$, demonstrating how processing conditions and interface engineering influence grain development, surface characteristics, and optical response. Thickness-dependent studies reveal a steady progression from densely nucleated nanocrystalline films to well-faceted surfaces, accompanied by lateral grain expansion and improved ordering during extended growth at a fixed temperature. As substrate temperature increases, enhanced surface mobility promotes grain coarsening and phase refinement, accompanied by higher growth rates and the resulting increase in film thickness. Notably, continuous, and phase-pure diamond films are achievable even at low temperature (480°C), highlighting the compatibility of this approach with device processing schemes constrained by limited thermal budgets.

See the Supplementary Material for additional experimental data and analysis supporting the main results. This includes control growth studies of diamond films on sapphire and Si substrates, comparative structural and optical characterization, and a detailed evaluation of the influence of $SiO_2$ and $SiN_x$ dielectric interlayers on diamond growth on β-$Ga_2O_3$. Control experiments on sapphire show grain size increasing from ~50 nm to ~202 nm with thickness (58 to 824 nm), accompanied by an improvement in $sp^3$-phase purity from ~97.3% to ~98.5%. Similarly, diamond films grown on Si exhibit large grain sizes (~246 nm) and high $sp^3$-phase purity (~98.9%) at ~814 nm thickness, confirming dense, well-coalesced films. For β-$Ga_2O_3$ substrates, diamond films grown on $SiO_2$ and $SiN_x$ interlayers under identical conditions exhibit comparable grain sizes (~126.6 nm vs. ~106.5 nm) and phase purity (~97.1% vs. ~96.9%), indicating only marginal influence of interlayer type.

**Acknowledgement**

The authors acknowledge the funding support from the National Science Foundation (NSF) under award numbers ECCS 2532898 and ECCS 2501623 and from the Air Force Research Laboratory Regional Network- Mid-Atlantic Hub.

**Data Availability**

The data that support the findings of this study are available from the corresponding author upon reasonable request.

**Conflict of Interest**

The authors have no conflicts to disclose.

**Figure 1**

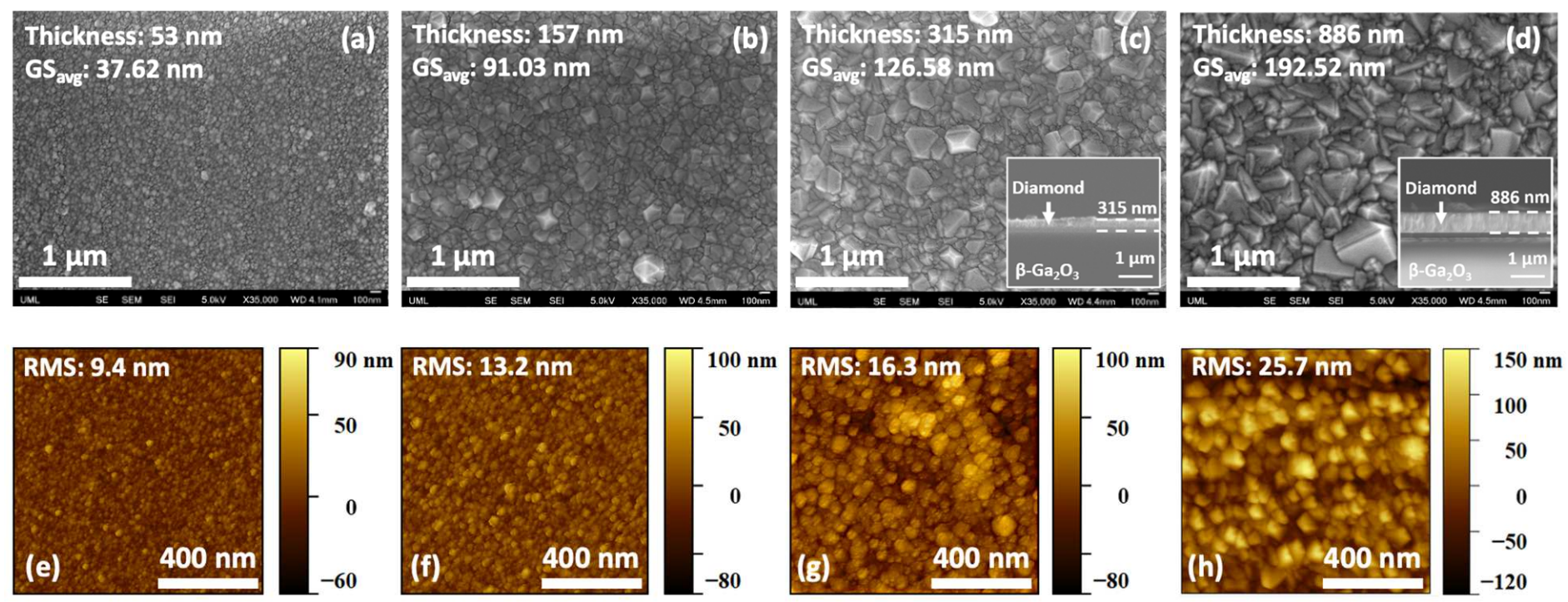

**Figure 1.** Surface morphology and roughness evolution of diamond films grown at 800 °C for 10, 30, 60 and 150 minutes: (a-c) SEM images and (d-f) corresponding AFM scans showing increasing grain size, faceting and RMS roughness.

**Figure 2**

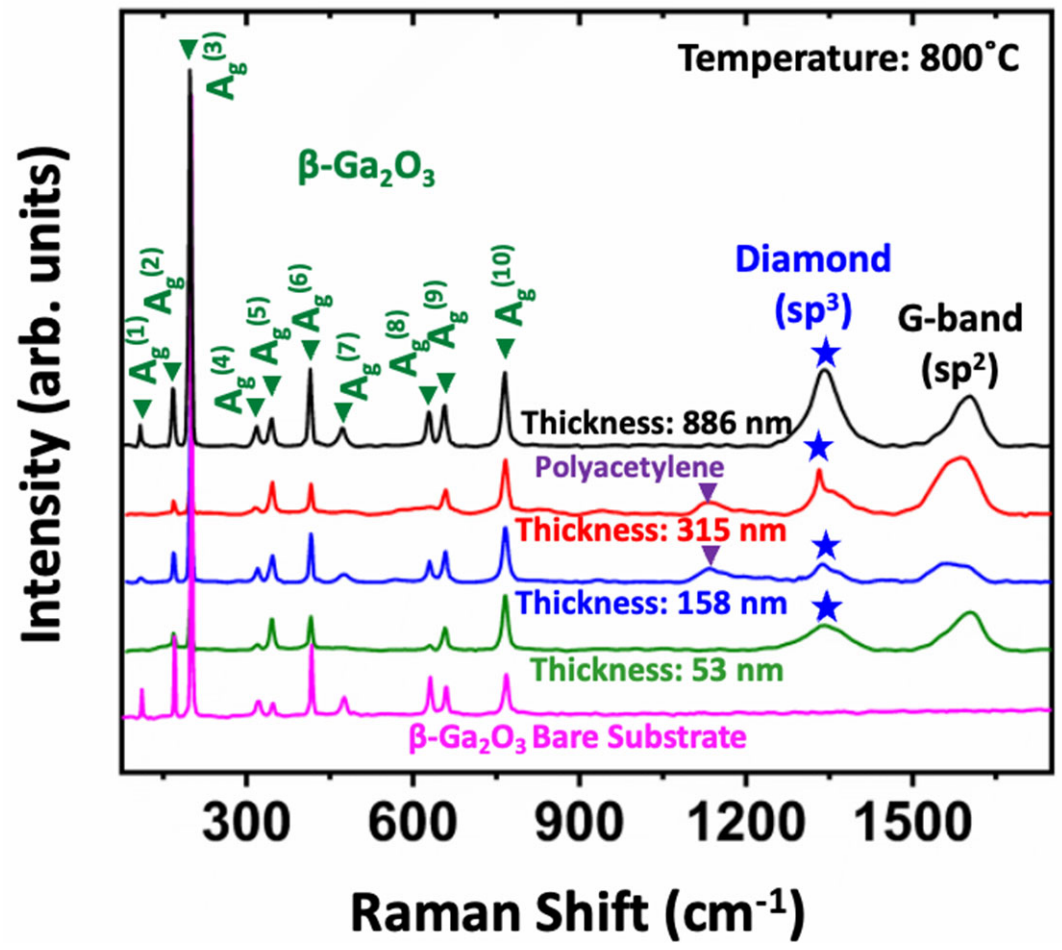

**Figure 2.** Raman spectra of diamond films grown at 800 °C, showing intensification of the sp$^3$ diamond peak with increasing film thickness.

**Figure 3**

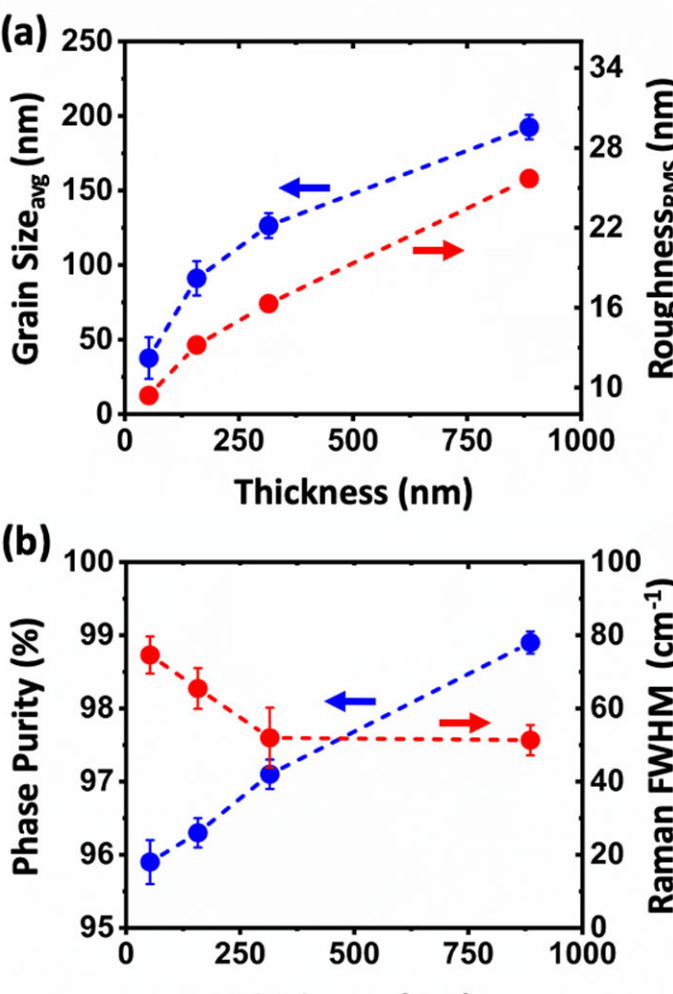

**Figure 3.** Grain size, surface roughness, Raman FWHM, and $sp^3$-phase purity as a function of diamond film thickness grown at 800 °C.

**Figure 4**

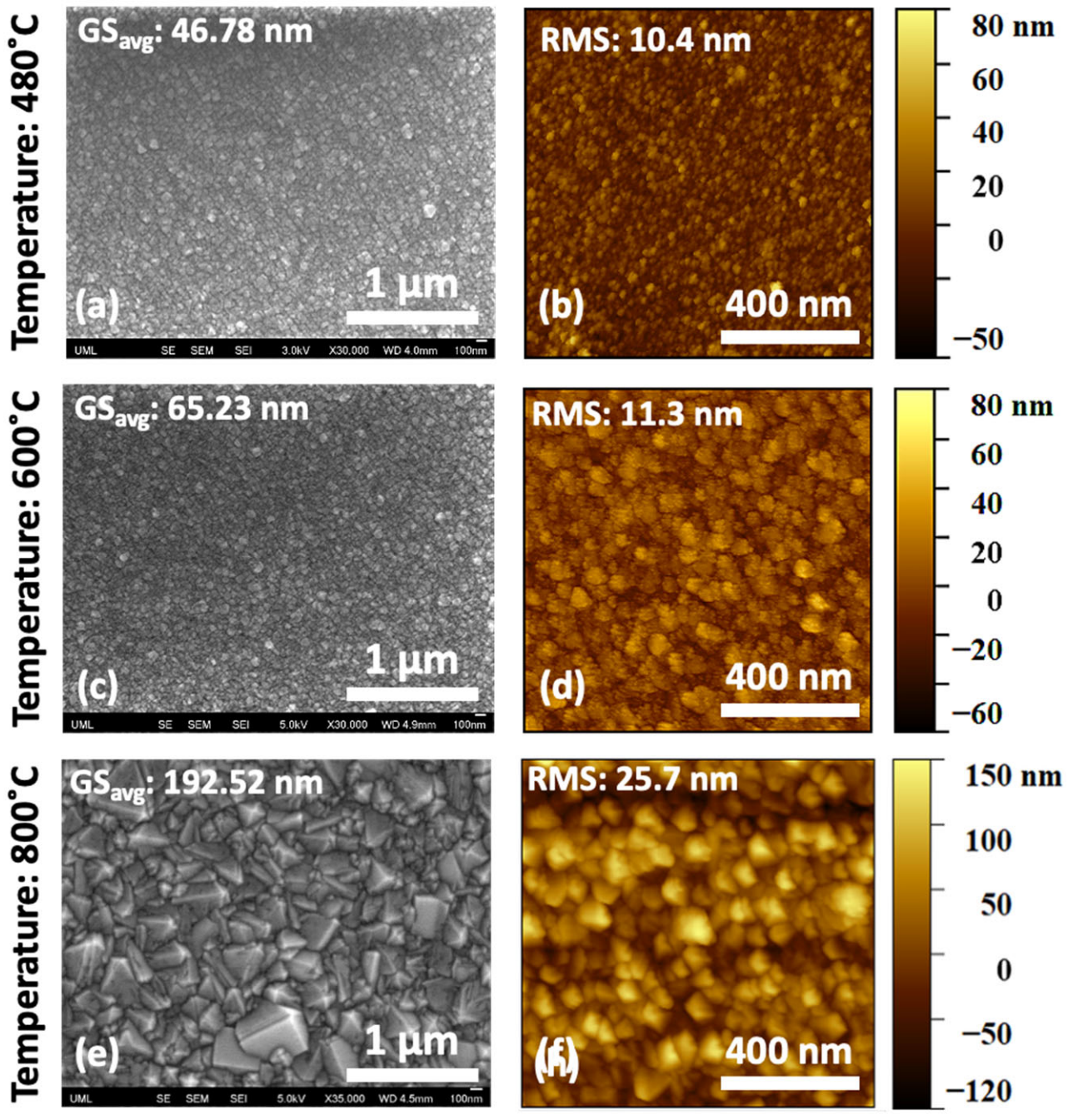

**Figure 4.** (a,c,e) SEM and (b,d,f) AFM images showing temperature-dependent evolution of surface morphology, grain size, and roughness of diamond films grown at 480 °C, 600 °C, and 800 °C.

**Figure 5**

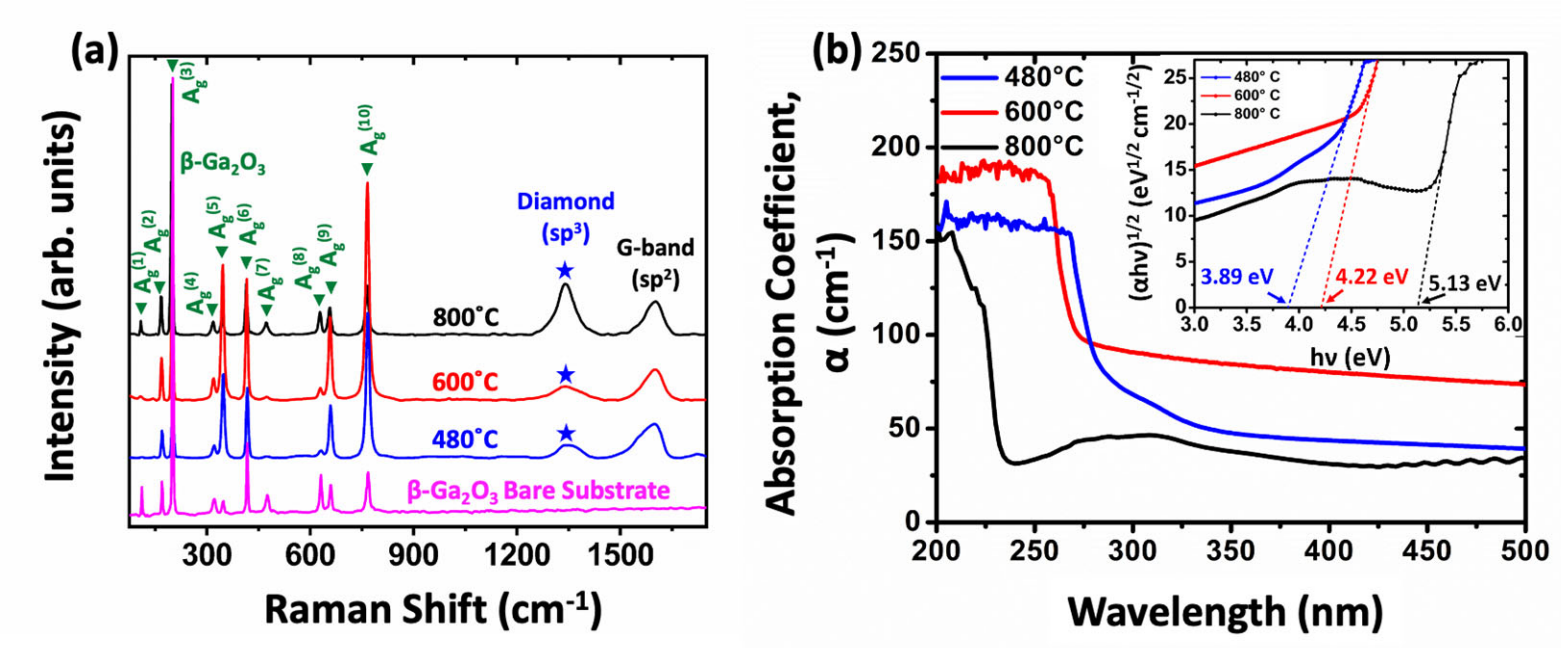

**Figure 5.** (a) Raman and (b) optical absorption spectra of diamond films grown at 480 °C, 600 °C, and 800 °C.

**Figure 6**

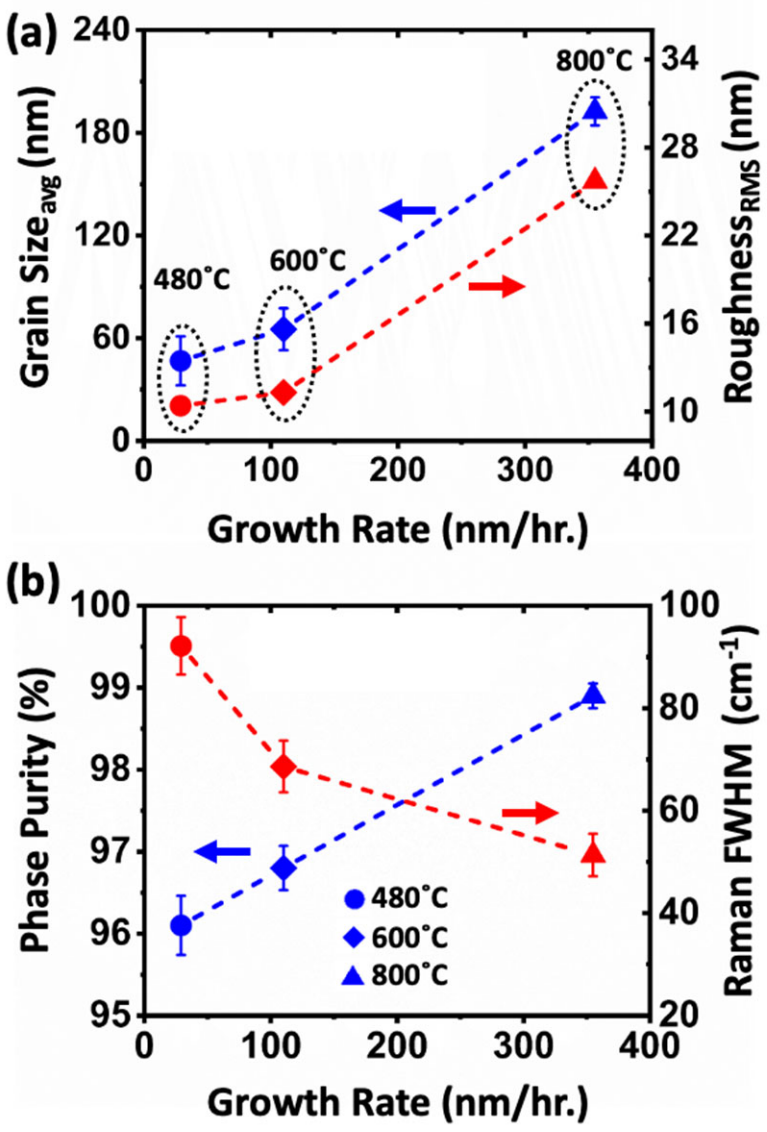

**Figure 6.** (a) Average grain size and surface roughness, and (b) Raman FWHM and $sp^3$-phase purity of diamond films as a function of growth rate achieved at substrate temperatures of 480 °C, 600 °C, and 800 °C.

**Figure 7**

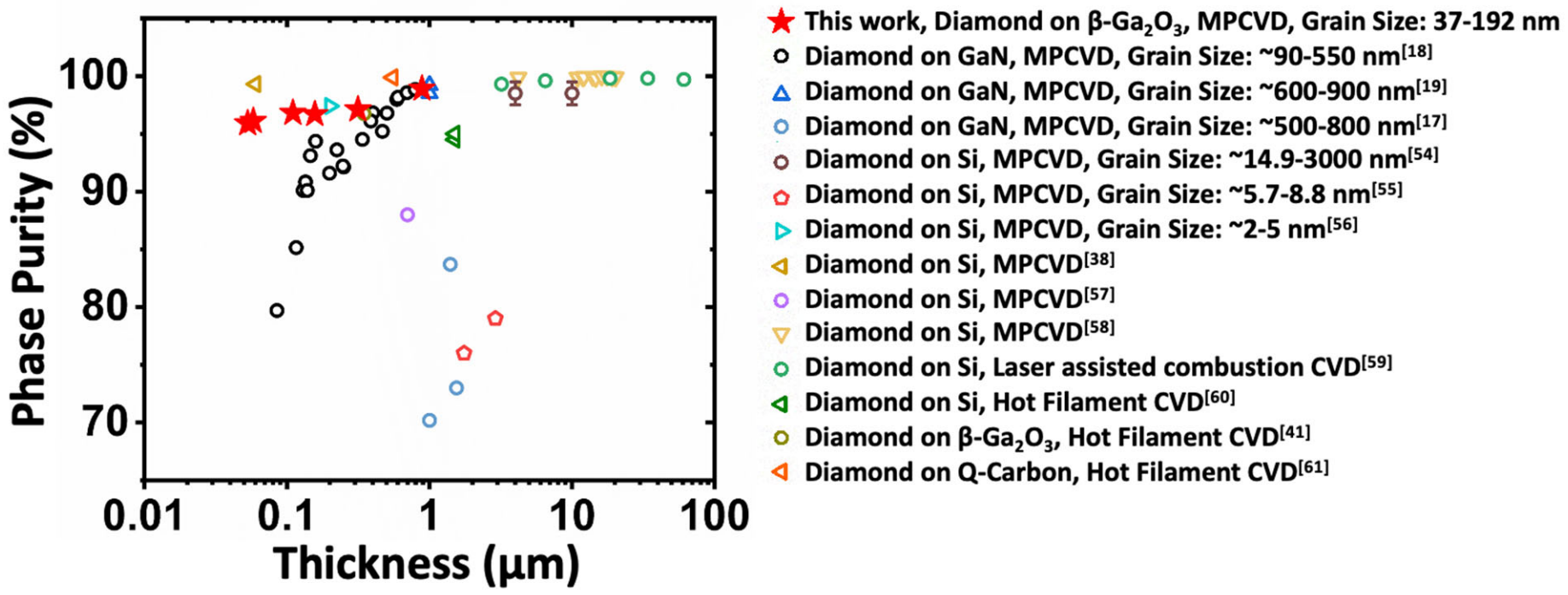

**Figure 7.** Benchmark comparison of $sp^3$-phase purity as a function of diamond film thickness for the present work and representative literature reports of polycrystalline diamond grown on foreign substrates using various CVD techniques.

# Supplementary Material

## Microwave plasma CVD growth and characterization of polycrystalline diamond films on β-$Ga_2O_3$

Saleh Ahmed Khan[1], Stephen Margiotta[1], Ahmed Ibreljic[1], Anhar Bhuiyan[1, a)]

[1]*Department of Electrical and Computer Engineering, University of Massachusetts Lowell, MA 01854, USA*

[a)] Corresponding author Email: anhar_bhuiyan@uml.edu

### I. Diamond Growth on Sapphire Substrates:

Control growth experiments were performed on c-plane sapphire substrates under MPCVD conditions comparable to those used for β-$Ga_2O_3$ (50 Torr, 1% $CH_4/H_2$, polymer-assisted nanodiamond seeding). Raman spectra [Supplementary Fig. S1(a)] confirm the formation of diamond with characteristic $sp^3$ features near 1332 $cm^{-1}$. Two films with thicknesses of approximately 58 nm and 824 nm were grown under these conditions. The 58 nm film exhibits an average grain size of ~50 nm and an $sp^3$-phase purity of ~97.3%, whereas the 824 nm film shows substantially larger grains (~202 nm) and an improved phase purity of ~98.5%. Cross-sectional imaging verifies uniform film coverage and good adhesion to the sapphire surface. These observations are consistent with progressive microstructural refinement during continued MPCVD deposition and provide a useful cross-substrate reference, indicating that the $sp^3$-phase fractions and grain coarsening behavior observed on β-$Ga_2O_3$ follow the same trend, where increased film thickness is associated with improved $sp^3$-phase purity.

**Table S1.** Diamond films deposited on sapphire substrates.

**Alt Text:** Summary of diamond films grown on sapphire showing that increasing film thickness leads to larger grain size and higher sp3 phase purity, indicating improved crystallinity with continued growth.

| Sample | Substrate | Diamond Thickness (nm) | Grain Size (nm) | Phase Purity (%) |
|---|---|---|---|---|
| 1 | $SiO_2$/Sapphire | 824 | 202 | 98.5 |
| 2 | $SiO_2$/Sapphire | 58 | 50 | 97.3 |

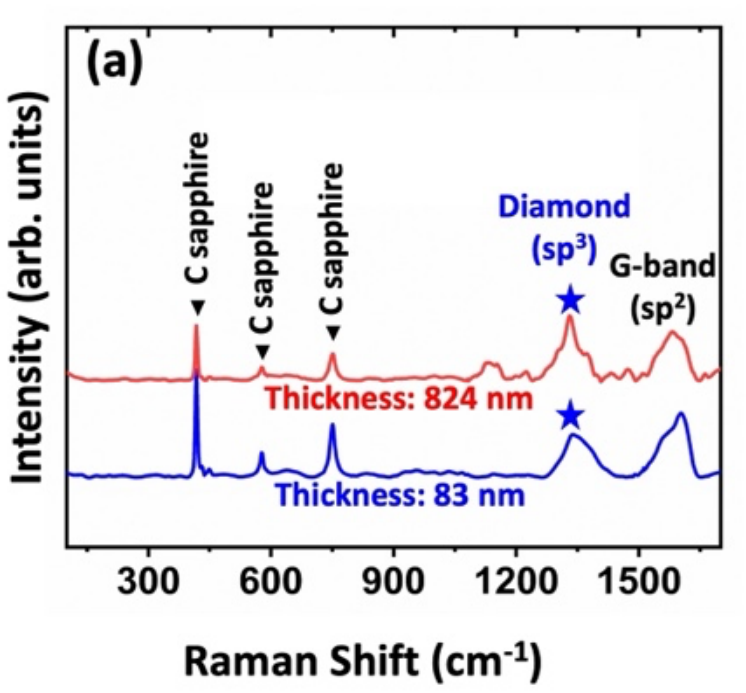

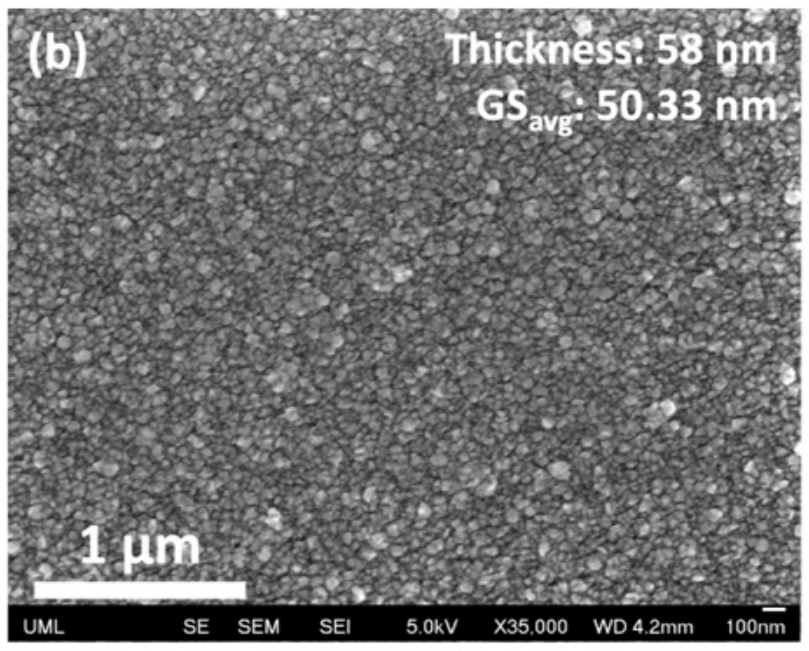

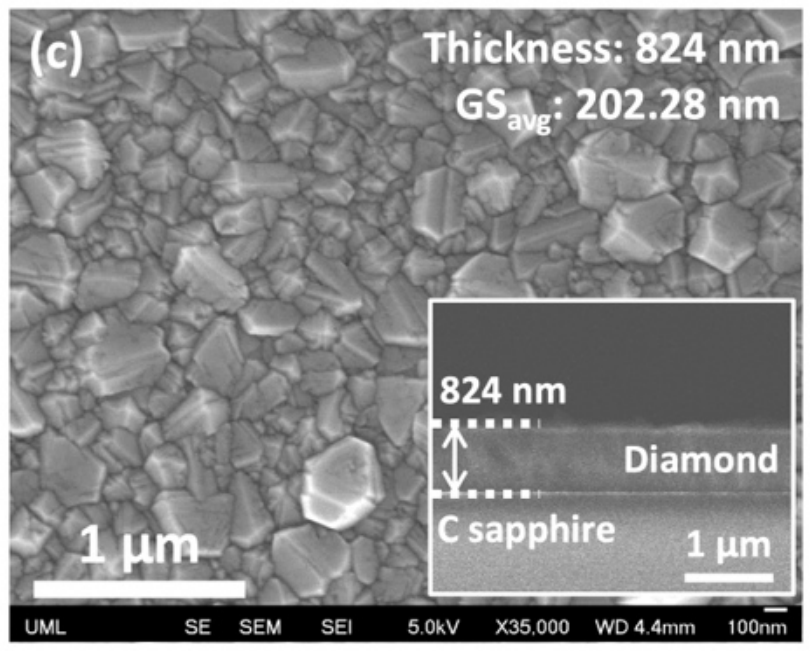

**Figure S1.** Diamond growth on sapphire substrates. (a) Room-temperature Raman spectra of diamond films grown on sapphire, showing the characteristic $sp^3$ diamond peak near 1332 $cm^{-1}$. (b) Top-view SEM image of a ~58 nm thick film exhibiting fine-grained morphology ($GS_{avg} \approx 50$ nm). (c) Top-view and cross-sectional SEM images of a thicker (~824 nm) film showing significant grain coarsening ($GS_{avg} \approx 202$ nm) and continuous film coverage on sapphire.

**Alt Text:** Diamond films grown on sapphire showing thickness-dependent grain growth. Grain size increases and crystallinity improves with increasing film thickness, consistent with progressive microstructural refinement during MPCVD growth.

## II. Diamond Growth on Si Substrates:

Control diamond growth experiments were also carried out on Si substrates under comparable MPCVD and polymer-assisted nanodiamond seeding conditions (50 Torr, 1% $CH_4/H_2$) for a deposition duration of 150 minutes. The Raman spectrum [Supplementary Fig. S2(a)] exhibits a well-defined $sp^3$ diamond peak near 1332 $cm^{-1}$, corresponding to a $sp^3$-phase purity of 98.9%. Top-view SEM imaging [Supplementary Fig. S2(b)] reveals a dense, faceted polycrystalline morphology with an average grain size of ~246 nm following extended deposition. Cross-sectional SEM analysis [Supplementary Fig. S2(c)] confirms continuous film coverage with a thickness of ~814 nm and uniform adhesion to the underlying silicon substrate.

**Table S2.** Diamond films deposited on Si substrates.

**Alt Text:** Summary of diamond films grown on Si substrates showing large grain size and high sp3 phase purity after extended growth, consistent with dense and well-coalesced polycrystalline films.

| Sample | Substrate | Diamond Thickness (nm) | Grain Size (nm) | Phase Purity (%) |
|---|---|---|---|---|
| 1 | Si | 814 | 246 | 98.9 |

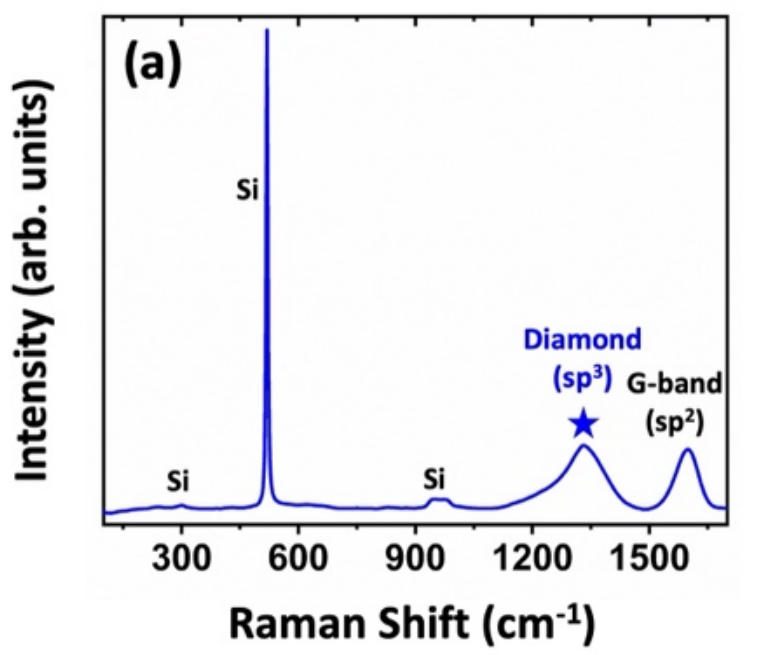

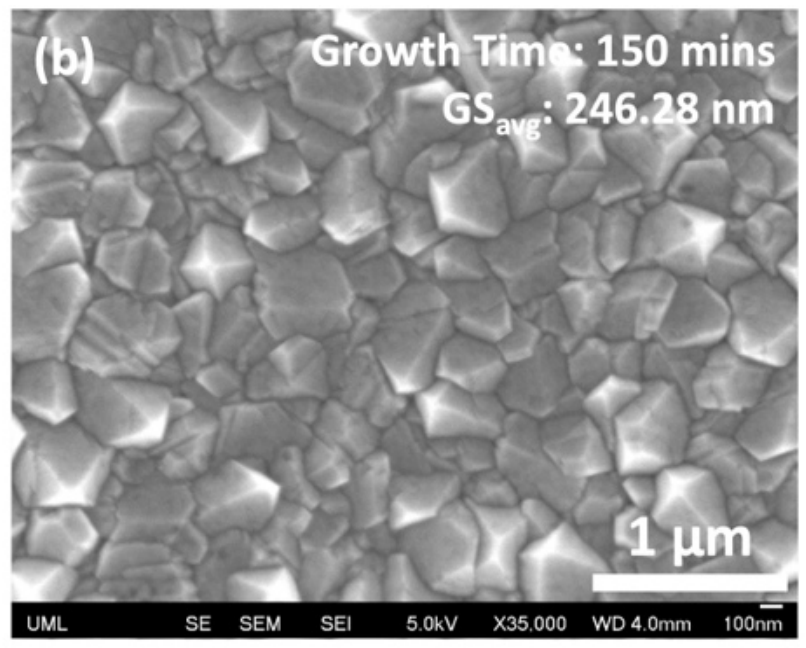

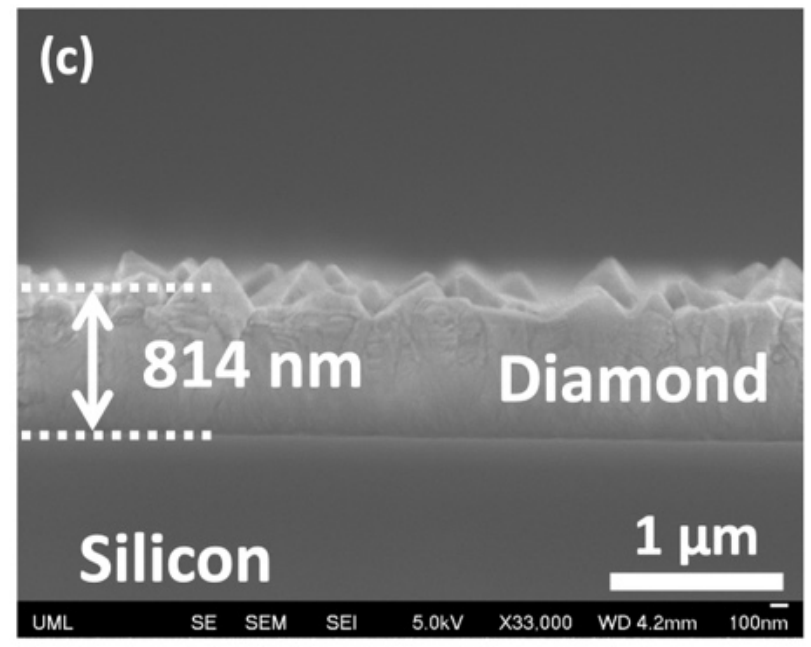

**Figure S2.** Diamond growth on Si substrates. (a) Room-temperature Raman spectrum showing a prominent $sp^3$ diamond peak near 1332 $cm^{-1}$. (b) Top-view SEM image of the ~814 nm thick film grown for 150 minutes, exhibiting well-developed faceted grains ($GS_{avg}$ ≈ 246 nm). (c) Cross-sectional SEM image confirming uniform film thickness (~814 nm) and continuous diamond coverage on the Si substrate.

**Alt Text:** Diamond films grown on Si substrate showing dense and continuous coverage. Large faceted grains and high sp3 phase purity are observed, indicating effective coalescence and improved film quality.

## III. Influence of Dielectric Interlayers on Diamond Growth on β-$Ga_2O_3$:

To assess the influence of dielectric interlayers on diamond film quality, Figure S3 compares the structural and optical properties of diamond films grown under identical MPCVD conditions on $SiO_2$- and $SiN_x$-coated β-$Ga_2O_3$ substrates. Both interlayers were deposited to a thickness of 50 nm via PECVD at 300 °C and 1 Torr using a plasma power of 20 W. The $SiO_2$ films were deposited using an $N_2O/SiH_4$ chemistry, yielding a refractive index of 1.47, consistent with near-stoichiometric oxide [1, 2]. $SiN_x$ films were deposited using an $N_2/NH_3/SiH_4$ chemistry, resulting in a refractive index of 1.91. This value lies within the typical range reported for PECVD silicon nitride films ($n \approx 1.8 - 2.1$), where the refractive index depends strongly on deposition conditions and film composition, particularly the Si/N ratio [3-5]. PECVD $SiN_x$ films deposited near 300 °C are generally non-stoichiometric and contain bonded hydrogen (Si–H and N–H), with reported N/Si ratios of ~1.2 - 1.35 and hydrogen contents of ~10 - 30 at.% depending on growth conditions [3-5]. Diamond growth was then performed under identical MPCVD conditions, including a substrate temperature of 800 °C, pressure of 50 Torr, $CH_4/H_2$ ratio of 1%, and growth duration of 60 minutes. SEM micrographs reveal that the film grown on $SiO_2$ exhibits an average lateral grain size of 126.6 nm, while the $SiN_x$-buffered counterpart shows an average grain size of 106.5 nm. The modest difference in grain size suggests that the interlayer may exert a limited influence on surface energetics or early-stage nucleation dynamics, but the overall microstructural evolution remains comparable under the same growth conditions. In addition, the continuous and conformal surface morphology without observable discontinuities or delamination suggests robust film integrity; however, atomic-resolution interface-sensitive characterization (e.g., scanning transmission electron microscopy combined with electron energy loss spectroscopy (STEM-EELS)) is required for definitive assessment of interfacial structure, bonding, and chemical composition, including potential interfacial reactions or diffusion at the $Ga_2O_3$-interlayer-diamond interface. The corresponding phase purity, estimated from Raman analysis, is 97.1% for the $SiO_2$ sample and 96.9% for the $SiN_x$ sample. The Raman spectra in Figure S3(a) for both samples exhibit diamond peaks near 1332 $cm^{-1}$ and similar secondary features attributed to polyacetylene and $sp^2$-bonded carbon. While slight differences in FWHM are observed, both films demonstrate comparable crystalline quality and successful diamond phase formation. The presence of $\beta$-$Ga_2O_3$ substrate phonon modes in both spectra confirms the transparency of the diamond layers and preservation of substrate structural integrity during growth. Figure S3(b) shows the optical absorption spectra and extracted Tauc plots. The film on $SiO_2$ exhibits an effective optical absorption edge of 5.13 eV, while the $SiN_x$-based film shows a comparable value of 5.11 eV. The small difference between these values falls within experimental uncertainty, indicating that the dielectric interlayer has only a marginal impact on the intrinsic optical transition. Although only marginal differences are observed between $SiO_2$ and $SiN_x$ interlayers under identical growth conditions, subtle variations in surface chemistry and interfacial energetics may influence early-stage nucleation and subsequent microstructural evolution. Differences in surface hydroxylation, charge distribution, and electrostatic interaction with the polymer-assisted nanodiamond layer could affect initial seed attachment and local nucleation density [6-8]. Variations in surface energy between oxide and nitride layers [9, 10] may also influence island wetting behavior and grain expansion kinetics during continued growth. Under hydrogen-rich plasma conditions, both interlayers act as chemically stable diffusion barriers that mitigate direct plasma interaction with β-$Ga_2O_3$ [11, 12]. Further interfacial characterization would be required to quantify these effects definitively.

**Table S3.** Structural and optical properties of diamond films grown on 50 nm $SiO_2$- and $SiN_x$-coated β-$Ga_2O_3$ substrates under identical MPCVD conditions (800 °C, 50 Torr, $CH_4/H_2$ = 1%, 60 min).

**Alt Text:** Comparison of diamond films grown on $SiO_2$- and $SiN_x$-coated β-$Ga_2O_3$ under identical conditions.

| Sample | Interlayer | Diamond Thickness (nm) | Grain Size (nm) | Phase Purity (%) |
|---|---|---|---|---|
| 1 | $SiO_2$ | 315 | 126.6 | 97.1 |
| 2 | $SiN_x$ | 315 | 106.5 | 96.9 |

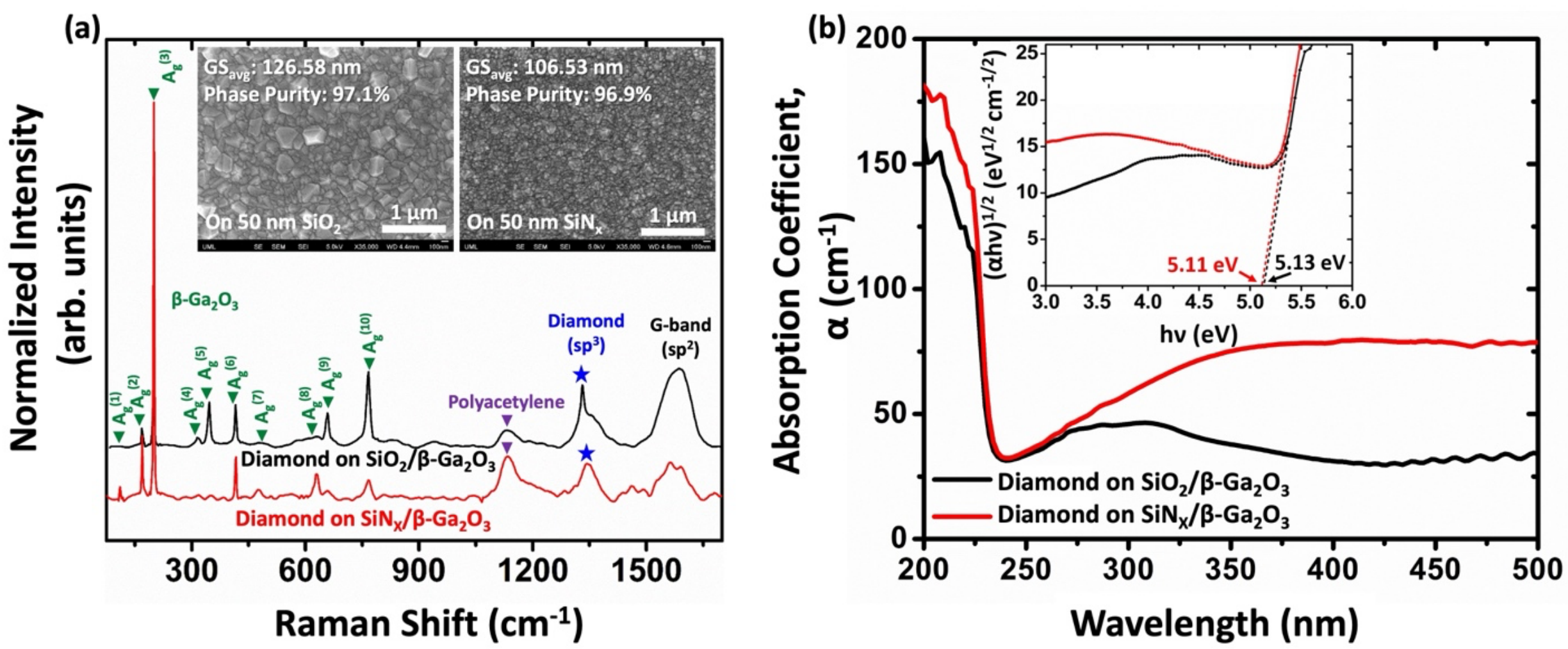

**Figure S3.** (a) Raman spectra and SEM images (insets) comparing grain size and phase purity of diamond films grown on $SiO_2$- and $SiN_x$-coated β-$Ga_2O_3$; (b) corresponding optical absorption spectra with extracted effective optical absorption edge.

**Alt Text:** Comparison of diamond films grown on $SiO_2$- and $SiN_x$-coated β-$Ga_2O_3$ under identical conditions.